\documentclass{article}
\usepackage[margin=1in]{geometry}
\usepackage[utf8]{inputenc}
\usepackage{graphicx}
\usepackage{xypic}
\usepackage{multicol}
\usepackage{titling}
\usepackage{xcolor}
\definecolor{Sage}{rgb}{0.53,0.64,0.49}
\definecolor{Navy}{rgb}{0.22,0.27,0.66}
\definecolor{Teal}{rgb}{0.13,0.49,0.44}

\usepackage{hyperref}

\hypersetup{
    colorlinks=true,
    linkcolor=Navy,
    filecolor=black,      
    urlcolor=Sage,
    citecolor=Teal,
    pdftitle={Towards a More Reasonable Semantic Web},
    pdfpagemode=FullScreen,
    }
    
\title{Towards A More Reasonable Semantic Web
}
\author{Vleer Doing, Rijkswaterstaat, {\sf vleer.doing@rws.nl} \\
Ryan Wisnesky, Conexus AI, {\sf ryan@conexus.com} } 

\begin{document}

\maketitle

\begin{abstract}
We aim to accelerate the original vision of the semantic web by revisiting design decisions that have defined the semantic web up until now. We propose a shift in direction that more broadly embraces existing data infrastructure by reconsidering the semantic web’s logical foundations. We argue to shift attention away from description logic, which has so far underpinned the semantic web, to a different fragment of first-order logic. We argue, using examples from the (geo)spatial domain, that by doing so, the semantic web can be approached as a traditional data migration and integration problem at a massive scale. That way, a huge amount of existing tools and theories can be deployed to the semantic web's benefit, and the original vision of ontology as shared abstraction be reinvigorated.
\end{abstract}

\tableofcontents

\pagebreak

\section{Introduction}

The semantic web envisions an ecosystem of seamless interaction enabled by ontologies, which are formally defined as logical theories \cite{gruber1995toward}, by associating the names of entities in the universe of discourse (e.g., classes, relations, functions, or other objects) with human-readable text describing what the names mean and with axioms that constrain the interpretation and use of these terms. 


Informally, this vision means that data needs to be augmented with additional structure ("semantics" and/or "metadata") so that it can be processed automatically by machines, potentially yielding data that is not explicitly present in the original data, and facilitating data exchange between different actors. With this goal in mind, standards such as RDF, RDFS, and OWL have been developed and much effort has been put into publishing semantics and axiomatizations for different domains. However at scale, implementation has been more difficult than anticipated and even flagship projects such as DBpedia have resulted in a large number of logical inconsistencies when reasoning is attempted within the Semantic Web framework \cite{paulheim2015serving}.

Under the moniker of Linked Data, the Semantic Web's  effort has now been more modestly focused on exposing data using URIs, RDF and SPARQL with little commitment to semantics \cite{glimm2012owl}. Ten years into Linked Data there are unresolved challenges towards arriving at a machine-readable and decentralized Web of data \cite{polleres2020more}. Most projects, including community projects like Wikidata, rarely re-use terms from other ontologies and most structured data is exchanged using basic central vocabularies (e.g. \url{https://schema.org}). Important relationships such as part-whole have been extensively studied in several domains but for reasons that we will discuss later, it is e.g. impossible to axiomatize proper parthood as strict order in the Semantic Web, let alone representing interesting distinctions between different mereologies \cite{keet2012representing}. 

In this paper we argue that to rejuvenate the original vision we must reconsider the semantic web’s foundations by shifting attention away from description logic~\cite{dl}(DL), which has so far underpinned the semantic web, to a different fragment of first-order logic~\cite{enderton} (FOL): the language of existential Horn clauses ~\cite{ehc}, which, under many different names (``regular logic'' (RL)~\cite{rl}, ``lifting problems''~\cite{SPIVAK_2013}, ``embedded dependencies''~\cite{fagin} (EDs), datalogE~\cite{datalogE}, etc), has traditionally been used for data migration and integration, in many different domains, for over a hundred years\footnote{The ``Dedekind–MacNeille completion'' of the rational numbers into the real numbers can be viewed as an instance of data migration of the kind we propose, an operation which returns in one of our case studies in this paper.}. 

By shifting focus, we can connect the semantic web community to the automated theorem proving community~\cite{tptp1}, the database theory community~\cite{alice}, the category theory community~\cite{awodey}, the type theory community~\cite{term}, and many other communities that have studied this fragment and developed tools for it, and thereby increase by an order of magnitude the tools available for use in the semantic web. 

We maintain backward compatibility with current semantic web standards since both DL and RL are fragments of FOL: their formulae can be translated in FOL \cite{flugel2021fowl} and the formats we discuss later.

\subsection{Outline and Contributions}

 In section~\ref{prelim}, we define the relational data model, the RDF data model, and provide other mathematical definitions; in that section we also define FOL, DL, and RL, describe how the semantic web deviates from FOL, and relate some of their properties to challenges in (geo)spatial information management where semantic web and its enhancements fail to add good value for complexity.  In section~\ref{case}  we explain, using a series of case studies, how existing technologies based on FOL and/or RL can help us to achieve many of the semantic web's original goals in a way that's both more approachable for the average programmer, and more useful for working with complex, heterogeneous data, than the OWL / DL commonly used in semantic web applications today.  In section~\ref{proposal} we describe our proposal in detail and describe a FOL interchange format already in active use by thousands of people and hundreds of automated reasoning engines, the TPTP format~\cite{tptp1}.  We conclude in section~\ref{conc} by providing a concrete plan of action for leveraging existing investments in semantic web in this new FOL-centric world for (geo)spatial information management specifically. 

\section{Definitions and Discussion}
\label{prelim}
In this section we define the various data models and languages referred to in this paper and discuss some of their properties.

\subsection{The Relational Model}

Let $R$ be a set, whose elements we call {\it relation names}, and let $arity : R \to \mathcal{N}$ be a function assigning each relation name in $R$ to a natural number (integer $\geq 0$), called that relation's {\it arity}.  The pair $(R, arity)$ is called a {\it relational signature}.  Let $Dom$ be a set, called the {\it domain}, whose elements we call {\it values}.  An $n$-ary {\it relation}  (a.k. ``table'') is defined as a subset of the $n$-ary Cartesian product $Dom \times \ldots \times Dom$.  Let $I(r)$ be a function assigning each relation name $r$ of $R$ to an $arity(r)$-ary relation.  The pair $(Dom, I)$ is called a {\it relational database} on relational signature $(R,arity)$ with domain $Dom$.  In practice and in theory, $Dom$ will often be a structured set, for example, including null values, dynamically typed values, error values, URLs, ``blank nodes'', etc, and we may wish to consider more sophisticated type systems, a point we defer for now.  

We often extend a relational signature $(R, arity)$ with ``intensional'' function symbols (in contrast to ``extensional'' relation symbols), whose meanings are not given by any database but are fixed in advance, for example, $+ : Dom , Dom \to Dom$ implementing addition, etc.  (Here, $+$ has $arity$ 3.)   Such function symbols are often called ``user-defined'' in the practice of data management.  Function symbols of arity 0 are called ``constant symbols''.    When we have relation symbols and function symbols, or the distinction is not relevant, we speak of {\it signatures} without the qualifier ``relational''.  Outside of database theory, a database in the sense above is often called a ``structure'' or an ``algebra''~\cite{term}.  A traditional introduction to database theory is~\cite{alice}.

\subsubsection{First-order Logic}

Given a signature $(R, arity)$, we may form the associated first-order logic (FOl) as follows.  Define a {\it term} to be either a variable (e.g. $x$), or an $n$-ary function name applied to $n$ terms (e.g., $+(x,-(z,y)$), usually written infix as $x+(z-y))$.  Define a {\it formula} to be an $n$-ary relation name applied to $n$ terms (e.g., ${\sf Father}(x,y)$) (sometimes called a membership predicate), or an equality of two terms $t_1,t_2$ (written $t_1=t_2$) (sometimes called an equality predicate), or the negation of a formula $p$ (written $\neg p$), or the conjunction (written $p \wedge q$) or disjunction (written $p \vee q$) of two formulae $p$ and $q$, or the universal (written $\forall x, p$) or existential quantification (written $\exists x, p$) of a formula $p$.  

Every fully-quantified formula (``sentence'') $p$ evaluates to true or false in a database $I$ in the usual way; when $p$ evaluates to true in $I$, we write $I \models p$ and say that $I$ is a model of $P$.  Associated to the first-order logic on signature $(R, arity)$ is a binary relation between formulas called {\it entailment}, where we write $p \vdash q$ to indicate that $q$ is entailed by $p$, that is, that assuming $p$ we may deduce $q$ using the usual axioms of first-order logic, which we do not write here~\cite{alice}, focusing instead on relational algebra.  We note for completeness however that entailment is important because $p \vdash q$ iff $I \models p$ implies $I \models q$; i.e., entailment axiomatizes model-theoretic consequence.  A traditional introduction to FOL is~\cite{enderton}.

In this paper, we define a {\it schema} to be a signature along with a set of fully-quantified formulae (so-called ``sentences''), also called a {\it theory} in FOL phrasing.  Hence, a relational database satisfying a schema is the same thing as a first-order model satisfying a theory; Database theory also considers deductive databases~\cite{deductive} such as RDF/OWL databases, which are theories in particular forms, so that a deductive database satisfying a schema is the same thing as a theory being entailed by another theory; where traditional model theory and database theory diverge is in their treatment of null values and a requirement that databases be representable on a computer, a distinction we gloss over in this paper. Our terminology is not entirely universal: while for example an SQL schema is understood to include its data integrity constraints as a logical theory, in other data models, such as JSON, a JSON schema is not usually understood to include any data integrity constraints, at least not without annotations.  

\subsubsection{Relational Algebra}

By a celebrated theorem of Codd~\cite{codd}, relational algebra corresponds exactly to a particular fragment of first-order logic, the so-called ``domain independent'' fragment.  Intuitively, it's easy to see how a ``set-comprehension'' through some formulae, such as $\{ (x,z) \ | \ {\sf Father}(x,y) \wedge {\sf Father}(y,z) \}$, can be understood as a query, e.g., finding all grandchildren and their grandparents, which can be interpreted as a database join.  However, some  formulae, when understood as set comprehensions, such as $\{ (x,z) \ | \ \neg {\sf Father}(x,y) \wedge \neg {\sf Father}(y,z) \}$ depend on the underlying domain of the database (in this case, the ``non-fathers'').  If we rule out formulae that depend on the underlying domain, and only depend on the tables, then the fragment of first-order logic that remains is called ``relational calculus'' and corresponds exactly to relational algebra in expressive power.  It turns out that it is undecidable when a formula is domain independent, although in practice simple syntactic conditions (such as ensuring all quantification is ``bounded'') ensure domain independence for all queries of interest.

\subsubsection{Existential Horn Clauses and the Chase}

In this section we will pay special attention to a particular fragment of first-order logic, regular logic, because it is the largest fragment of first-order logic that admits a kind of ``model completion'' operation, called ``the chase''~\cite{chase} that allows us to take databases and uniquely alter them to according to new theories, a key operation in data exchange in integration.  This operation has been independently discovered in the GIS domain; see our case study on merging geographic hierarchies.  Roughly speaking a first-order formula is an existential horn clause when it has the form 
$$
\forall x_0, \ldots x_n, \phi \rightarrow \exists y_0, \ldots, y_m, \psi
$$
where $\phi$ and $\psi$ are (possibly empty) conjunctions of equalities and relation-memberships, and regular logic is defined to be that fragment of first-order logic where all formulae are existential Horn clauses.  Conceptually, if such a formula does not hold in a database, then to complete the database, we look for domain values $x_0, \ldots, x_n$ that match $\phi$ and then create new domain values (blank nodes, in the case of RDF) corresponding to $y_0, \ldots, y_n$ and add new tuples (or equate existing domain values) to the existing database so as to make $\psi$ true.  

A common example of the chase procedure is computing transitive closure.  Assume a binary relation named ${\sf Ancestor}$.  The formula:
$$
\forall x y z, {\sf Ancestor}(x,y) \wedge {\sf Ancestor}(y, z)  \to {\sf Ancestor}(x, z)
$$
adds one level of ancestry to the database when interpreted as a relational algebra query. When we chase with this formula, we iterate this operation until it reaches a fixed point.  E.g.
$$
{\sf Ancestor}({\sf alice},{\sf bob}) \wedge {\sf Ancestor}({\sf bob},{\sf charlie})  \wedge {\sf Ancestor}({\sf charlie},{\sf david})
$$
chases to
$$
{\sf Ancestor}({\sf alice},{\sf bob}) \wedge {\sf Ancestor}({\sf bob},{\sf charlie})  \wedge {\sf Ancestor}({\sf charlie},{\sf david}) \wedge $$
$${\sf Ancestor}({\sf alice},{\sf charlie}) \wedge {\sf Ancestor}({\sf alice},{\sf david}) \wedge   {\sf Ancestor}({\sf bob},{\sf david})
$$

 Whether or not the chase eventually stops is undecidable, but when it does stop, the result is unique up to ``homomorphic equivalence'' (the size of the resulting model will not be uniquely defined, but all resulting models will be mappable to each other).  The chase procedure generalizes the ``initial model construction'' of datalog and equational logic~\cite{term} to include existential quantifiers to the right of implications, which requires some additional technical complexity (labelled nulls, Skolemization) we elide here.

\subsubsection{Nested Relations, GraphQL, and LINQ}

When nested relations are required, the relational calculus and algebra can be extended into the so-called nested relational calculus and nested relational algebra~\cite{wong}, an extension that may not actually add expressive power in most data-centric settings.  For example, XML nested to $N$ levels can be represented using relations nested to $N$ levels, which can be represented using $N$ relations connected by ``foreign keys''.  In fact, this is how many XML databases actually work~\cite{10.1145/2588555.2612186}.  

The GraphQL language~\url{https://graphql.org} is essentially the type theory for nested relational algebra, defining table names and their columns and nested tables, making its name something of a misnomer.  We view the use of nested schemas vs flat schemas as a design choice, and as such are neutral to pro GraphQL. 

Language integrated queries~\cite{linq} (LINQ) are a technique for embedding the nested relational calculus into a higher-order programming language, for example, Microsoft C\# or Haskell.  This embedding is mathematically dual to the embedding of the nested relational calculus into flat relational calculus, in a sense that can be made precise using the language of category theory~\cite{corel}.  We a pro-LINQ: whenever we are functional programming, LINQ is our preferred way to interact with relational databases.  However, we don't see LINQ as being directly related to the semantic web.

\subsection{The Graph Model}

In this paper we take the view that graph databases are special cases of relational databases. We define a graph data model in part to illustrate how FOL works, and also because RDF data sets can be interpreted as certain kind graphs.  

In particular, a {\it directed, labelled multi-graph} is any database on the relational schema with three unary relation names, ${\sf node}, {\sf edge}, {\sf label}$ and four binary relation names, ${\sf src}, {\sf dst}, {\sf nlabel}, {\sf elabel}$, obeying axioms that $src$ and $dst$ are functions from ${\sf edge}$ to ${\sf node}$ and that ${\sf nlabel}$ is a function from ${\sf node}$ to ${\sf label}$ and that ${\sf elabel}$ is a function from ${\sf edge}$ to ${\sf label}$, which we can write in FOL (in fact, in RL) as:

$$
\forall x y, {\sf src}(x,y) \to {\sf edge}(x) \wedge node(x) \ \ \ \ 
\forall x y, {\sf dst}(x,y) \to {\sf edge}(x) \wedge {\sf node}(x) 
$$$$
\forall x y z, {\sf src}(x,y) \wedge {\sf src}(x,z) \to y=z \ \ \ \ 
\forall x y z, {\sf dst}(x,y) \wedge {\sf dst}(x,z) \to y=z 
$$$$
\forall x, {\sf edge}(x) \to \exists y, {\sf src}(x,y) \ \ \ \ 
\forall x, {\sf edge}(x) \to \exists y, {\sf dst}(x,y) 
$$$$
\forall x y, {\sf nlabel}(x,y) \to {\sf node}(x) \wedge {\sf label}(y) \ \ \ \ 
\forall x y, {\sf elabel}(x,y) \to {\sf edge}(x) \wedge {\sf label}(x) 
$$$$
\forall x y z, {\sf nlabel}(x,y) \wedge {\sf nlabel}(x,z) \to y=z \ \ \ \ 
\forall x y z, {\sf elabel}(x,y) \wedge {\sf elabel}(x,z) \to y=z 
$$$$
\forall x, {\sf edge}(x) \to \exists y, {\sf elabel}(x,y) \ \ \ \ 
\forall x, {\sf node}(x) \to \exists y, {\sf nlabel}(x,y) 
$$
For example:

$$
    \xymatrix{
    {\sf Alice} \ar@/^1.0pc/[rr]^{\sf mom} \ar@/^1.0pc/[rrr]^{\sf dad} & {\sf Bob} \ar@/_1.0pc/[r]_{\sf mom} \ar@/_1.0pc/[rr]_{\sf dad} & {\sf Megan} & {\sf Dan}
    }
$$

becomes the first-order model with
$$
Dom := \{ {\sf n_1, n_2, n_3, n_4, e_1, e_2, e_3, e_4, Alice, Bob, Megan, Dan, mom, dad} \}
$$
satisfying

$$
{\sf node}({\sf n_1}), {\sf node}({\sf n_2}), {\sf node}({\sf n_3}), {\sf node}({\sf n_4}),
{\sf edge}({\sf e_1}), {\sf edge}({\sf e_2}), {\sf edge}({\sf e_3}), {\sf edge}({\sf e_4}),
$$$$
{\sf label}({\sf Alice}), {\sf label}({\sf Bob}), {\sf label}({\sf Megan}), {\sf label}({\sf Dan}),
{\sf label}({\sf mom}), {\sf label}({\sf dad}),
$$$$
{\sf src}({\sf e_1}, {\sf n_1}), {\sf dst}({\sf e_1}, {\sf n_2}),
{\sf src}({\sf e_2}, {\sf n_1}), {\sf dst}({\sf e_2}, {\sf n_4}),
{\sf src}({\sf e_3}, {\sf n_2}), {\sf dst}({\sf e_3}, {\sf n_3}),
{\sf src}({\sf e_4}, {\sf n_2}), {\sf dst}({\sf e_4}, {\sf n_4}),
$$
$$
{\sf nlabel}({\sf n_1}, {\sf Alice}), {\sf nlabel}({\sf n_2}, {\sf Bob}),
{\sf nlabel}({\sf n_3}, {\sf Megan}), {\sf nlabel}({\sf n_4}, {\sf Dan})
$$$$
{\sf elabel}({\sf n_1}, {\sf mom}), {\sf elabel}({\sf n_2}, {\sf dad}),
{\sf elAbel}({\sf n_3}, {\sf mom}), {\sf elabel}({\sf n_4}, {\sf dad})
$$
Or, written tabularly,
$$
\begin{tabular}{ | c | }
\multicolumn{1}{c}{{\sf node}} \\
 \hline ${\sf n_1}$ \\  
 \hline  ${\sf n_2}$ \\
 \hline  ${\sf n_3}$ \\ 
 \hline  ${\sf n_4}$ \\ \hline 
\end{tabular}
\ \ \ \ 
\begin{tabular}{ | c | }
\multicolumn{1}{c}{{\sf label}}  \\ 
 \hline  {\sf Alice} \\  
 \hline {\sf Bob} \\
 \hline {\sf Megan} \\ 
\hline  {\sf Dan} \\
 \hline {\sf mom} \\
 \hline {\sf dad} \\ \hline  
\end{tabular}
\ \ \ \ 
\begin{tabular}{ | c | c | }
\multicolumn{2}{c}{{\sf nlabel}}  \\ 
 \hline ${\sf n_1}$ & {\sf Alice} \\
 \hline ${\sf n_2}$ & {\sf Bob} \\
 \hline ${\sf n_3}$ & {\sf Megan} \\
 \hline ${\sf n_4}$ & {\sf Dan} \\ \hline 
\end{tabular}
\ \ \ \ 
\begin{tabular}{ | c | }
\multicolumn{1}{c}{{\sf edge}}  \\ 
 \hline  ${\sf e_1}$ \\  
 \hline  ${\sf e_2}$ \\
\hline  ${\sf e_3}$ \\ 
 \hline ${\sf e_4}$ \\ \hline 
\end{tabular}
\ \ \ \ 
\begin{tabular}{ | c | c | }
\multicolumn{2}{c}{{\sf src}}  \\ 
\hline  ${\sf e_1}$ & ${\sf n_1}$ \\ 
\hline  ${\sf e_2}$ & ${\sf n_1}$ \\
\hline  ${\sf e_3}$ & ${\sf n_2}$ \\
\hline  ${\sf e_4}$ & ${\sf n_2}$ \\ \hline 
\end{tabular}
\ \ \ \ 
\begin{tabular}{ | c | c | }
\multicolumn{2}{c}{{\sf dst}}  \\ 
\hline  ${\sf e_1}$ & ${\sf n_3}$  \\
\hline  ${\sf e_2}$ & ${\sf n_4}$ \\
\hline  ${\sf e_3}$ & ${\sf n_3}$ \\
\hline  ${\sf e_4}$ & ${\sf n_4}$ \\ \hline 
\end{tabular}
\ \ \ \ 
\begin{tabular}{ | c | c | }
\multicolumn{2}{c}{{\sf elabel}}  \\ 
 \hline ${\sf e_1}$ & {\sf mom} \\
 \hline ${\sf e_2}$ & {\sf dad} \\
 \hline ${\sf e_3}$ & {\sf mom} \\
 \hline ${\sf e_4}$ & {\sf dad} \\ \hline 
\end{tabular}
$$
In this way, many graph databases are representable as databases conforming to relational schemas. Of course, there are many kinds of graph, and as can be intuited from the above example, each can be represented using a different relational schema, lending credence to the argument that FOL (and RL) is a useful foundation for semantic interchange.  For example, if we desire uni-graphs instead of multi-graphs, we might add an axiom
$$
\forall x y z w, {\sf src}(x,z) \wedge {\sf src}(y,z) \wedge {\sf dst}(x,w) \wedge {\sf dst}(y,w) \to x = y
$$
and as such both uni-graphs and multi-graphs can co-exist within FOL (each graph would have its own node and edge relations, etc).

Sometimes, we may wish to consider a graph as an instance on a schema whose relation names are themselves drawn from graph data.  In this encoding, ${\sf mom}$ and ${\sf dad}$ become binary relation names and we add facts ${\sf mom(n_1,n_3),dad(n_1,d_4),mom(n_2,n_3),dad(n_2,n_4)}$ to our database.  We will see a similar encoding of description logic into first-order logic later, when we introduce RDF.

\subsubsection{NoSQL and Multi Model}

As we say in the section on graphs, any set of FOL axioms can define a data model, and in fact, the proliferation of data processing systems can be understood in exactly that way.  For example, NoSQL (schema-less) databases such as MongoDB can be thought of as relational databases made up entirely of functions (tables of key-value pairs) \cite{corel}, and the more recent ``multi-model'' systems such as mm-adt~\url{https://www.mm-adt.org} which have emerged and allow an application to store data in one data model and later another application to query the same data using a different data model via defining multi-model data views (\cite{liu2018multi}. 

We may ask why noSQL and multi-model databases exist if they are merely special cases of relational databases.  The answer can be found in the type of queries typically run against such databases. That is, queries against such databases tend to involve particular operations which are hard to expressed in SQL (for example: many graph queries require a recursive SQL queries which are hard to express). Hence, vendors focus on providing query languages with constructs that extend relational algebra, as well as providing additional performance enhancement compared to standard relational systems. This is an arms race however, and whether vendors have been successful is another matter. Some benchmarks show relational databases consistently outperforming ``graph-first'' databases  \cite{mhedhbi2021lsqb}, including typical RDF workloads \cite{ravat2020efficient}. The upcoming SQL:2023 standard will bring MATCH syntax for pattern matching and path-finding in ``property graphs'' similar to neo4j’s Cypher query language and is being implemented in at least one popular SQL implementation \cite{ten2022integrating}.

\subsection{The Semantic Web Model}

In this section we define RDF, RDFS, and OWL.  Because the RDF specification alone is 22 pages long and also makes reference to URI/L and XML standards we follow~\cite{blanknodes} in defining a simplification of RDF and related technologies at an abstract level, pointing out issues in their definitions as we go, particularly in a geospatial context.  For a retrospective on the semantic web, twenty years on, see~\cite{twoDecades}.

\subsubsection{RDFS, OWL and the URI-first approach}

We assume the existence of pairwise disjoint infinite sets U (URIs), L (literals) and B (blank nodes), where following custom~\cite{blanknodes} we write UB for the union of U and B, and similarly for other combinations.  A RDF triple is defined as a tuple $(s,p,o)$ in $UB \times U \times UBL$, where $s$ is called the subject, $p$ the predicate and $o$ the object.  An RDF database is defined as a set of RDF triples.  An RDF database is also called an RDF graph, although as we will see there are many ways to understand RDF as graphs.

We begin by noting how URIs permeate the definition of RDF: they can be found in $s$, and in $p$, and in $o$. This is unfortunate, in our opinion. One issue is that the meaning of URLs is ambiguous - is the URI just a name/string, serving as a kind of hierarchical namespace?  Should it be dereferenced in a web browser? How exactly does a URI differ from a URL? What if the referent changes over time? Will the referent be XML, or RDF, or something else?  What if one web actor's answer to these questions is different than another actor's?  To start to answer these questions, we begin with an example RDFS dataset:

$$({\sf example:dog_1}, {\sf rdf:type}, {\sf example:animal}), 
$$
$$
({\sf example:cat_1}, {\sf rdf:type}, {\sf example:cat}),
({\sf example:cat_1}, {\sf rdfs:subClassOf}, {\sf example:animal}), 
$$ $$
({\sf example:host}, {\sf rdfs:range},	{\sf example:animal}),
({\sf example:zoo_1}, {\sf example:host}, {\sf example:cat_2})$$

In the example above, and we can see how in RDF ${\sf host}$, conceptually a relation name, must be a URI, because only URIs are common to subjects and predicates.  Hence technically we must write ${\sf example:cat1}$ and ${\sf example:host}$ and so forth; we thus say that RDF forces a commitment to the ontology of URIs, as a user must now understand e.g., how URIs factor according to $:$, etc\footnote{We note that ironically, URL manipulation operations themselves (e.g., split under ``:'') are more naturally expressed as functions than triples.}.  We note also that we had to ``invent'' the word {\sf example}; we could just as easily have chosen {\sf scenario}, and consumers of such an RDF dataset may not know if this choice is meaningful: we believe that lightweight uses of URIs in the current form introduce more complexity via requiring arbitrary choices than is reclaimed by their ability to "uniquely" identify resources.

RDFS defines a controlled vocabulary using URIs, where this vocabulary is designed to be a ``vocabulary about vocabulary'' common across most RDF databases. One might say that RDFS and OWL are designed to solve exactly the problem RDF and SPARQL ignore, namely, providing the user with entailment/deduction. RDF graphs can be validated against RDFS specifications, providing some protection against the ``predicates are also subjects'' issues of plain RDF (namely, whatever protection users happen to axiomatize in any particular RDF graph). 

A subset of RDFS vocabulary is shown below.   \\

\hspace{-.3in}
\includegraphics[width=1.6in]{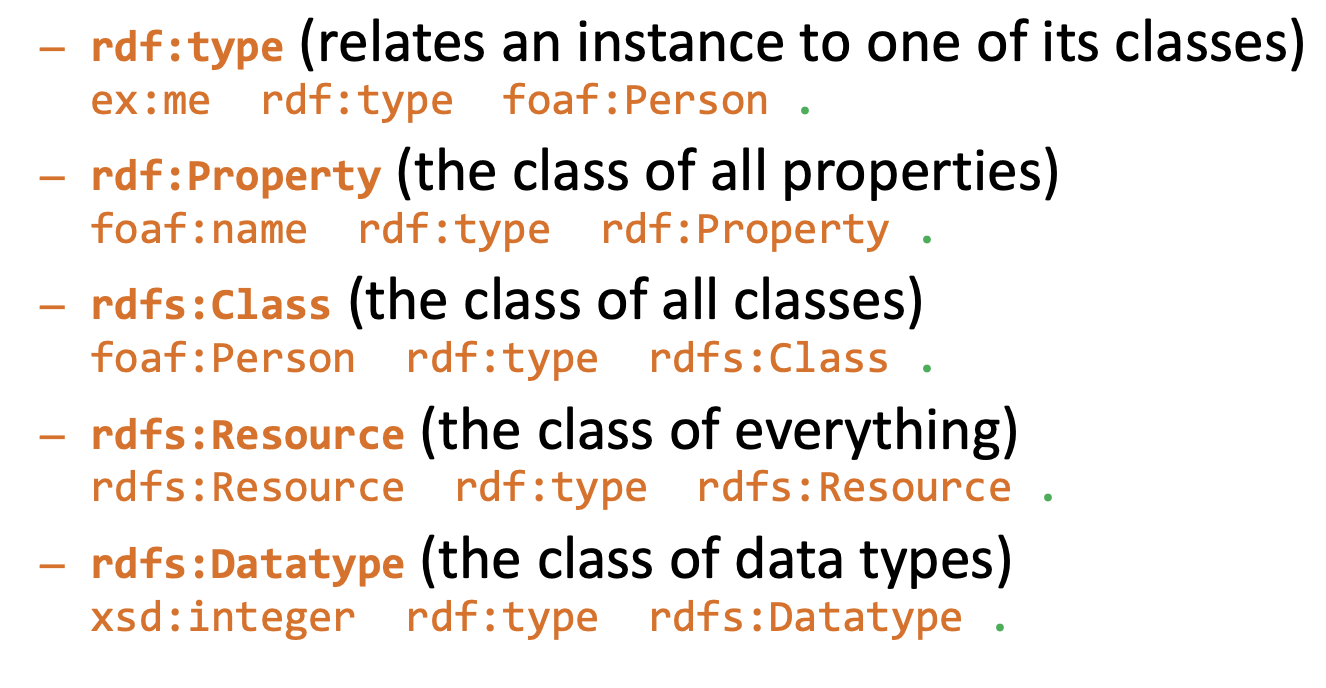}
\includegraphics[width=1.6in]{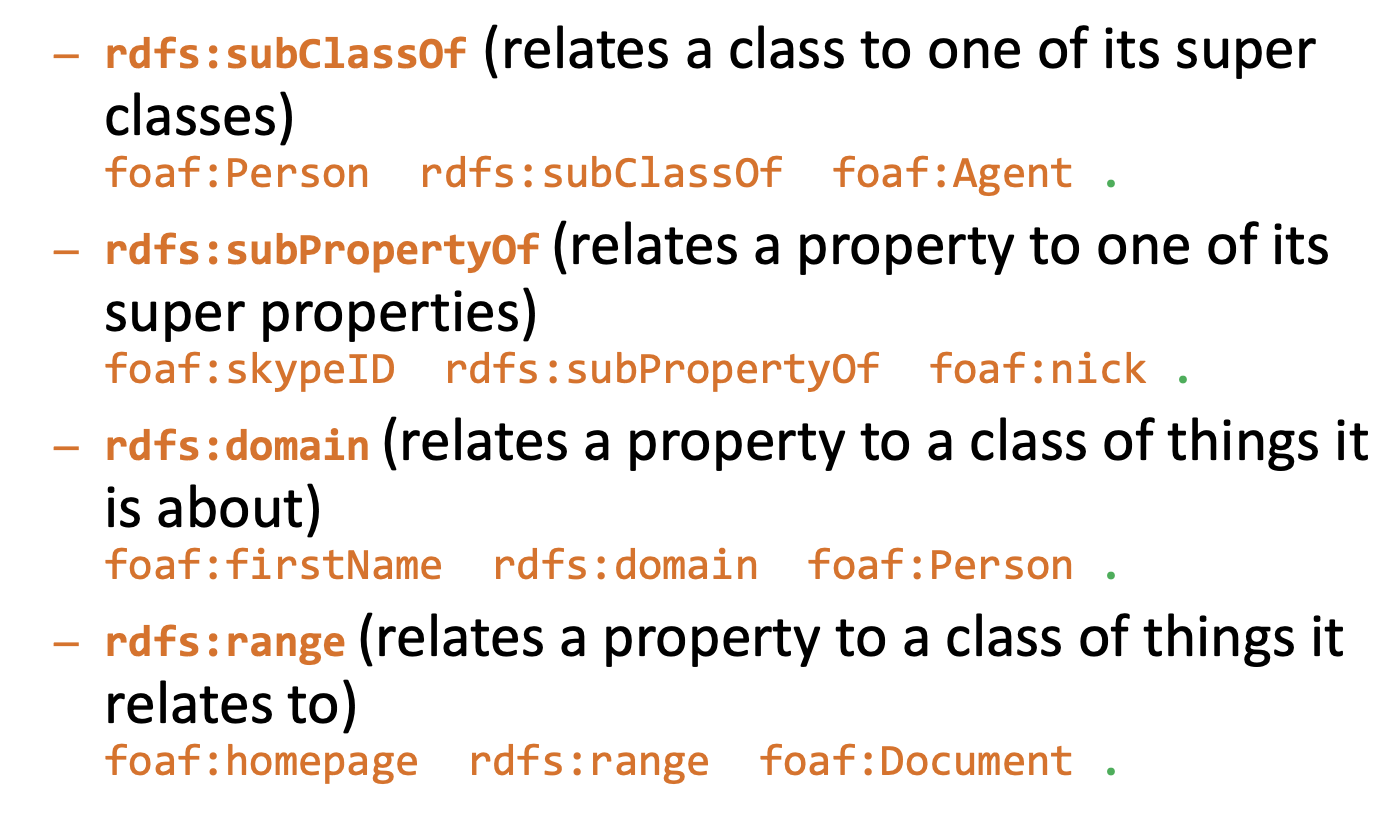}
\includegraphics[width=1.6in]{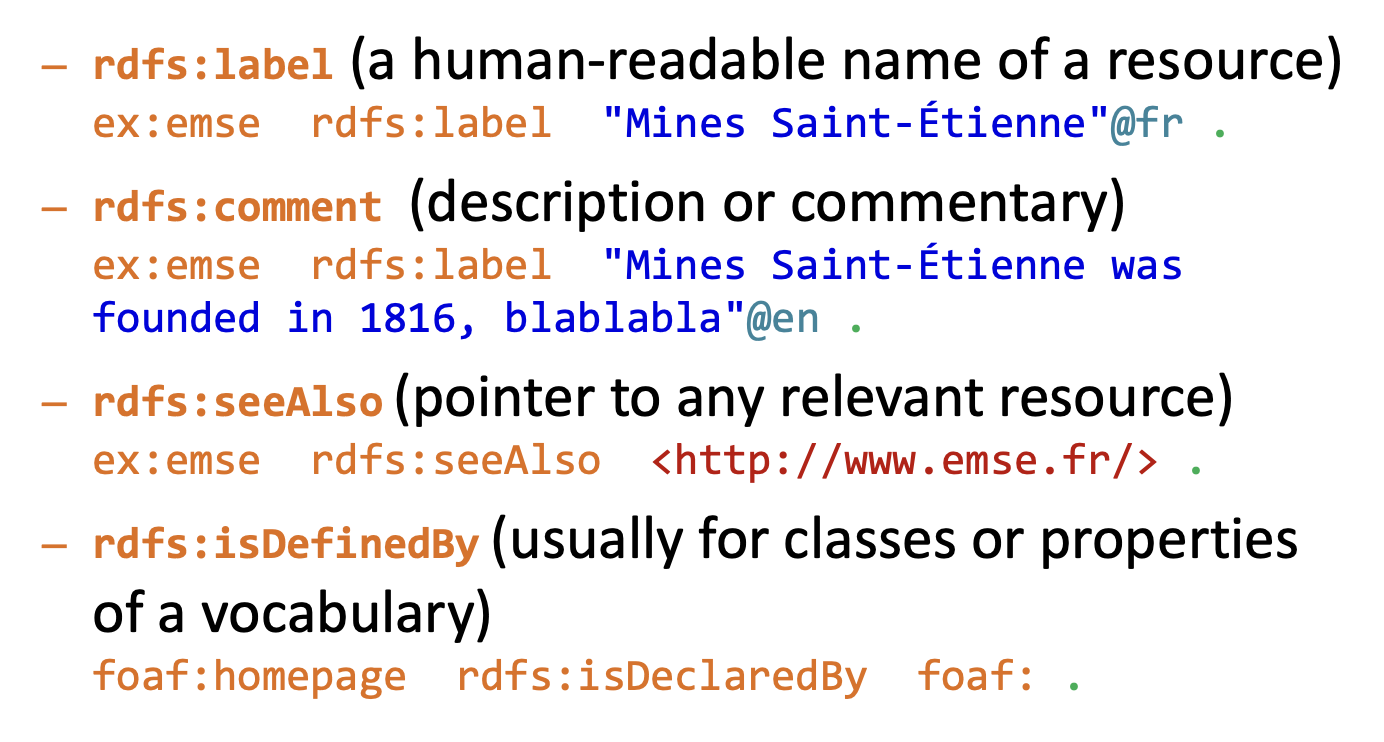}
\includegraphics[width=1.6in]{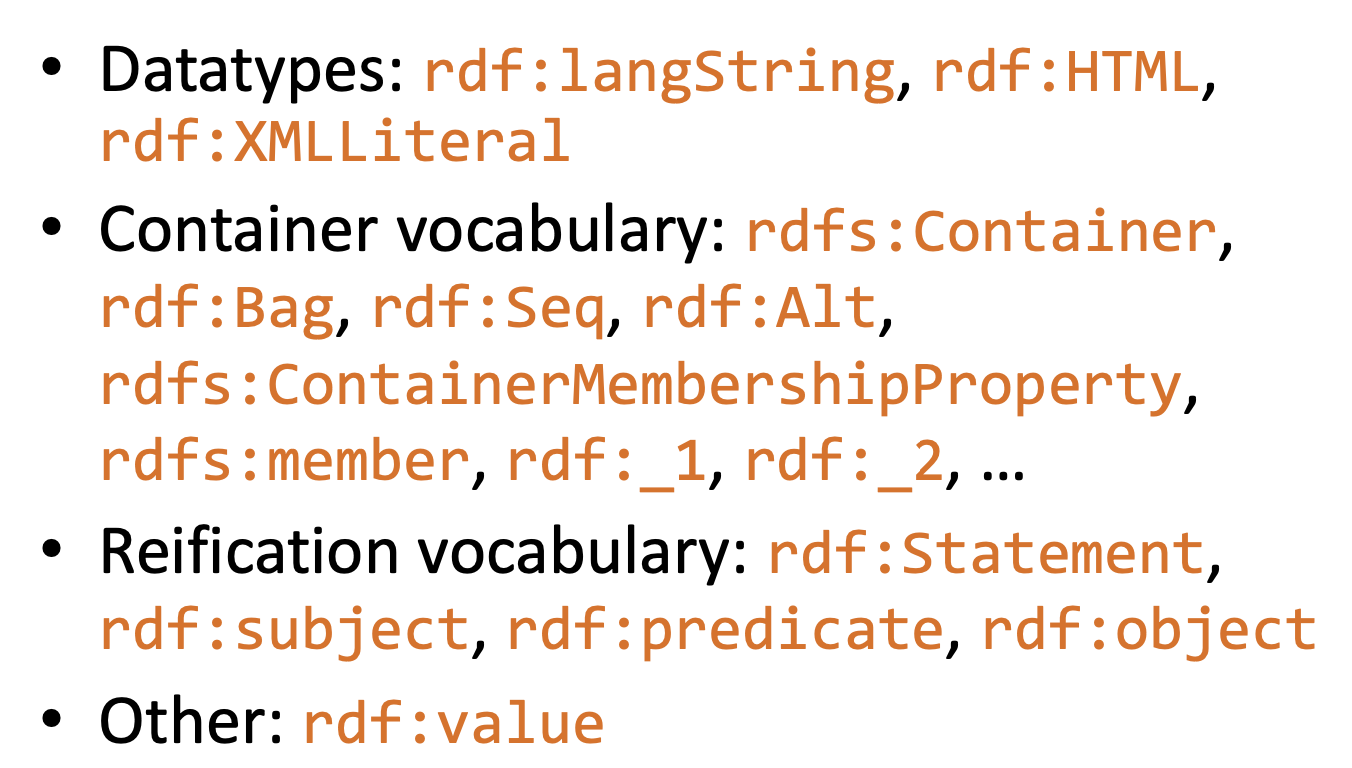}

A priori, there's no reason to expect or desire a schema language to be encodeable as an instance of the very data model it is providing the schema for\footnote{And it is notable when something like this does happen, for example, how relational conjunctive queries can be encoded as databases~\cite{alice}.}; Of course, one might say that part of the point of the semantic web is to force a commitment to it's own syntax of URI's, in which case, we suggest axiomatizing the semantics as well, for example an RDF predicate {\sf deref} such that: 
$$
\forall x:U, \ \exists y:B, \ {\sf deref}(x, y) \ \ \ \ \forall x y y' : U, \ {\sf deref}(x, y) \wedge {\sf deref}(x, y') \to y = y'
$$

When an RDF program is inferenced against it, it can be appended with triples of the form ${\sf deref}(u, s)$ where $s$ is the string literal obtained by actually dereferencing the URL $u$ on the web, hence providing an offline semantics and an online semantics that refines it.  If both reading and writing of URIs is desired, such as in \url{https://www.w3.org/TR/ldp-primer/}, then axiomatization becomes more complex, because we have to model a distributed, aysynchronous state-based system evolving over time, which can involve clock skew, race conditions, and the usual problems of distributed computation.  There are many formalisms for reasoning about such systems, which take us beyond the scope of this paper; here we remark briefly on two such systems.  The first, TLA+~\cite{tla}, has seen industrial use at Microsoft and LinkedIn, and is based on an extension of FOL known as ``temporal logic''. The second, concurrent separation logic~\cite{10.1145/2984450.2984457}, is based on an extension of FOL known as ``Hoare logic''.  They, and other similar logics, are useful for proving safety and liveness properties of such systems, such as deadlock-freeness, as well reasoning about data structure invariants, for example, proving that a concurrent implementation of a B-Tree is race-free.  


Despite the vocabulary of RDFS and OWL being common to most/all applications of the semantic web, many subtle bugs enter the Semantic Web through it due to its divergence from regular FOL. For example, one of the main mechanism to relate vocabulary that are supposed to be integrated are {\sf owl:sameAs} statements. In practice these lead to a tough problem, known in the Semantic Web community as the sameAs problem \cite{raad2019sameas}. It is the result from naively equating meaning even though semantics that actually vary depending on context; this problem occurs in e.g. enterprise-wide data integration, where words such as e.g. ``cost'' and ``risk'' often have different meanings and so are not {\sf sameAs} across departments. This {\sf sameAs} problem of context-depended meaning is prevalent in the (geo)spatial semantic web where geometric information is involved  \cite{beck2021evaluation}. Since ontologists who work on foundational issues typically prefer FOL, examples of such being DOLCE and BFO, there's the unfortunate situation that (geo)spatial domain ontology written in OWL are hard to verify against their associated upper level ontology and interoperability is hindered. The FOL preference can be traced back to issues within RDF(S) and OWL which we will discuss in the next section.

\begin{itemize}
    \item  One challenge comes from the asymmetry in the definition of RDF: the set $L$ only appears in $o$, and not in $s$. If instead an RDF triple was a member of $BL \times U \times BL$, then we could treat RDF URLs as identifiers for relation names, disjoint from a notion of literal that exists along side blank nodes $B$ (a kind of labelled null); in this case, there is a very natural encoding of an RDF graph (set of tuples) as a first-order theory with domain $L$, with blank nodes encoded by existential quantifiers; and if the RDF graph had no blank nodes, there would be a very natural encoding of the RDF graph as a first-order model with domain $L$. We believe that in practice, this scenario happens a lot, and this special ``normal form'' of RDF is key to the interplay between RDF and graph databases, because first-order models in this form can be read as encoding graphs in a way similar to the previous section. Let us call this {\it the RDF asymmetry problem}.

\item Further difficulties come from too much symmetry in the definition of an RDF triple: the fact that URLs can appear in all three subject/predicate/object positions means that RDF is in some sense higher-order, or self-referential, and not in a good way: formulae such as $X(X,X)$ that are ill-typed in FOL are allowed, where $X$ is any URI, because $(X,X,X)$ is a valid RDF triple.  As such, arbitrary RDF triples do not correspond to ``shallow embeddings'' of graphs in first-order logic unless users maintain a discipline which partitions URIs into distinct classes, one for predicates and another for individuals.  Such a discipline is in fact used in encodings of JSON into FOL.   Let us call this {\it the URL-subjects-and-predicates problems}.  

\item Additionally RDF has a very limited set of syntactic constructs: no other construct except for triples is allowed, neither in RDF(S), nor in OWL, or related languages. This implies that a technique known as “RDF reification” must be used to present N-ary relationships for N $>$ 3. For example, to encode an 5-ary tuple of a relation R, say, $R(a,b,c,d,e)$, requires five binary tuples and
a ``fresh'' identifier, say, $x$: $R_1(x, a), R_2(x, b), R_3(x, c), R_4(x, d), R_5(x, e)$. Then, to ``use'' the 5-tuple requires joining five relations together. 

This type of encoding makes representing Spatial data in RDF is non-trivial. Typical feature types such as Polygon or MultiLineString are decomposed resulting in a large total amount of statements where every separate data object has its own URI (the $x$ above). This increases the data size significantly \cite{stepien2022approach} and makes processing the grouped feature, e.g., perform calculations, index it, etc., difficult, as the feature first needs to be reassembled from its parts (the $R_i$ above) \cite{brodt2010deep}. Alternative notations and abbreviations such as WKT literals are suggested as a syntactic workaround but require (pre)processing outside of the Semantic Web. Let us call this the {\it everything is a triple problem.}

\item RDF literals are used to represent values such as strings, numbers and dates. The datatype abstraction used in RDF is compatible with XML Schema, and in fact, RDF re-uses many of the XML Schema built-in datatypes, and defines two additional ``non-normative'' datatypes, rdf:HTML and rdf:XMLLiteral. Our critique here is primarily that XML is an unsound and incomplete interchange format~\cite{xml}, and as such should never be used for anything, including RDF; practically any other type system than XML's would make RDF easier to understand and implement.  Let us call this the {\it XML type system problem}.

\item Despite its XML-based type system, RDF lacks any mechanism for adding computational (vs axiomatic) behavior, typically known as ``user-defined functions'', such as addition.  At best, one may, using the ``full'' fragment of OWL, axiomatize the natural numbers so that e.g. ``2 + 2'' is deductively equivalent to ``4''.  But then we lose the ability to lose OWL reasoners, which are not based on the full fragment of OWL;

Such a ``deductive database''~\cite{deductive} must also be queried in such a way as to respect the axioms of arithmetic, something that SPARQL cannot because of its issues with blank nodes, which we discuss next. 

Let us call this {\it the UDFs vs decidability problem}. 

\item In the previous section on first-order logic and relational algebra, we saw how we could use FOL as a query language.  A similar idea underlies the SPARQL query language, but with a caveat: SPARQL does not distinguish between U and B and L, treating all of them as though they were literals (all members of L).  This means that a SPARQL query can give different results depending on how blank nodes are used in an RDF database: if $I$ and $J$ are RDF databases containing labelled nulls, and $th_I$ and $th_J$ are the associated first-order theories satisfying $I$ and $J$, and $I$ and $J$ are logically equivalent, there may exist a SPARQL query $Q$ for which $Q(th_I)$ and $Q(th_J)$ are not logically equivalent.  In other words, SPARQL does not preserve logical deduction / entailment. Let us call this the {\it SPARL doesn't respect blank nodes} problem.

Fortunately, however, in practice, blank nodes are rare~\cite{blanknodes}, meaning that SPARQL's deviation from RDF's semantics doesn't manifest often.  Unfortunately, however, this deviation from RDF semantics means that an RDF dataset cannot both axiomatize arithmetic and be safely SPARQL-queryable at the same time, a fact which hinders interoperability.  

\end{itemize}

\subsubsection{Description Logic}
Description logic (DL)~\cite{dl} is a decidable fragment of FOL that is the basis for the OWL language, decidable meaning that there exists an algorithm that always returns true or false when asked if one formula logically entails another ($\vdash$).  The syntax for DL is written to the left below, and the translation $[]$ of DL into FOL is shown on the right below.
\begin{multicols}{2}
  \begin{tabular}{ l c l c l }
$C,D$ & ::= & {\sf A} & $\mid$ & atomic concept \\  
 &  & $\top$ & $\mid$ & universal concept \\  
 &  & $\bot$ & $\mid$ & empty concept \\  
 &  & $\neg C$ & $\mid$ & complement \\
 &  & $C \sqcup D$ & $\mid$ & union \\
 &  & $C \sqcap D$ & $\mid$ & intersection \\
 &  & $\exists R, C$ & $\mid$ & existential restriction \\
 &  & $\forall R, C$ & $\mid$ & universal restriction 
\end{tabular}
  
 $$
[A]_x := A(x) \ \ \ \ 
[\neg C]_x := \neg [C]_x$$ $$
[\forall R, C]_x := \exists y, R(x,y) \to [C]_y \ \ \ \ 
[C \sqcap D]_x = [C]_x \wedge [D]_x $$ $$
[\exists R, C]_x := \exists y, R(x,y) \wedge [C]_y \ \ \ \ 
[C \sqcup D]_x = [C]_x \vee [D]_x$$ $$
[C \sqsubseteq D] := \forall x, [C]_x \to [D]_x
$$
\end{multicols}

For example,
$$
[{\sf Animal} \sqcap \forall {\sf hasParent}, {\sf Donkey}]_x =
{\sf Animal}(x) \wedge \forall y, {\sf hasParent}(x, y) \to {\sf Donkey}(y) 
$$ $$
[{\sf Animal} \sqsubseteq {\sf LivingThing}] = \forall x, {\sf Animal}(x) \to {\sf LivingThing}(x)
$$
\begin{itemize}

\item Description Logic is somewhat awkward to use for the goals of the semantic web, at least when building the semantic web is understood as data integration, simply because many theories of common data models do not fall into the two-variable fragment of FOL; for example, the theory of multi-part foreign keys in a relational database.  Moreover, the decidability of DL implies that it cannot be used to express mathematical theories at or beyond basic arithmetic, limiting its applicability because arithmetic is useful when integrating web data.  We propose regular logic, with its formulae of existential Horn clauses, as better alternative.  Let us call this the {\it DL was designed for 90s computers problem}.

\item Individuals, classes and RDF-properties are all elements in the RDF domain, but class extensions are only implicitly defined by the rdf:type property.  This means for example that ``a subClassOf b'' cannot be understood as the straightforward existence of an injective function from a to b, where a and b are disjoint partitions of the domain, without performing an analysis of the rdf:type property.   Pushed to its extreme, because RDFS supports reflection on its own syntax -- it is defined in terms of classes and RDF-properties which are interpreted in the same way as other classes and RDF properties, and whose meaning can be extended by statements in the language -- it is possible to argue that RDFS in its full generality doesn't even posses a (direct) set-theoretic model, so that when we try to (shallowly) translate RDFS in its entirety into some FOL sub-language we meet Russell's paradox, which we've seen already in the form of the RDF triple $X(X,X)$.  Let us call this the {\it what is the meaning of RDFS type} problem.

\item OWL is an attempt to encode description logic into RDF.  However, because description logic is based on FOL and RDF is not, the encoding has a number of technical issues, first among them being that OWL has two distinct semantics: one based of first-order logic, that applies only to a fragment of RDF, and another based on RDF's 22-page full semantics, that applies to all of RDF, with a ``correspondence theorem'' stating that the two semantics agree when they overlap. So one does not technically use OWL, one uses OWL-direct or OWL-rdf, with theorem required to relate them.  Let us call this the {\it  OWL semantics problem}. 
 
 \item The key property of a description logic, decidability of entailment, does guarantee the termination of certain algorithms in finite time, but if such algorithms take, for example, doubly exponential time, what difference does termination make in practice, where heuristics must be used anyway?\footnote{To be fair, there are many fragments of description logic, many with very fast algorithms.}  We believe that, to the extent possible, existential Horn clauses form a better logic for the goals of the semantic web, because in such cases it is possible to uniquely ``repair'' (chase) models/databases to satisfy theories, something that cannot be done for description logic (because of the presence of disjunction).  For example, the DL formula:
  $$
 {\sf Actor} \sqcap {\sf USGovernor} \sqsubseteq {\sf Bodybuilder} \sqcup \neg {\sf Austrian}
 $$
 Becomes the FOL formula:
 $$
\forall x, {\sf Actor}(x) \wedge {\sf USGovernor}(x) \to {\sf Bodybuilder}(x) \vee \neg {\sf Austrian}(x)
 $$
 
Such a formula lacks ``repairs'' because if you have an actor and US governor who is neither a body builder nor not Australian, there is no canonical choice of which model the repair should be - do you make them a body builder, or do you make them non-Austrian, or both?  Logics with disjunction at best admit ``multi-repairs'', i.e., databases can be repaired into unique {\it sets} of databases that individually satisfy the given theory.  It is for this reason that relational data integration technology has traditionally favored the logic of existential Horn clauses over other, more expressive logic: RL posses ``certain answers'', tuples that must occur in all solutions, but DL does not.  The existence of certain answers allows us to meaningfully query the result of a repaired/chased database without having to consider how it was repaired/chased.

Of course, when one is not interested in data integration, but instead in curating a single database, then the description logic vs existential Horn clauses tradeoff is reversed - the additional modeling power of disjunction may be desirable.  Fortunately, because both description logic and existential Horn clauses are fragments of first order logic, we don't have to choose!

\item 
In logics with negation such as description logic/OWL, one must chose whether to interpret the absence of a fact (say, {\sf Alice} not appearing in the {\sf Person} table) as the presence of the negation of the fact (say, $\neg({\sf Person}({\sf Alice}))$.  FOL and DL do not do this, and thus make the ``open-world'' assumption.  This choice is considered controversial in the semantic web community, because certain recent extensions to RDF such as SHACL seem to require the opposite choice, the ``closed-world'' assumption.  Our first take on this ``controversy'' is that it was already settled in 1977 in a paper~\cite{closed}: you just pick which semantics you want, and convert closed-world query answering to open-world query answering, being careful about the exact conversion based on whether or not you have an extensional (tabular) or intensional (deductive) database.  We hold this result as an example of the power of traditional results in data integration to provide design guidance for the semantic web.  We quote the original paper here: 

\includegraphics[width=5.5in]{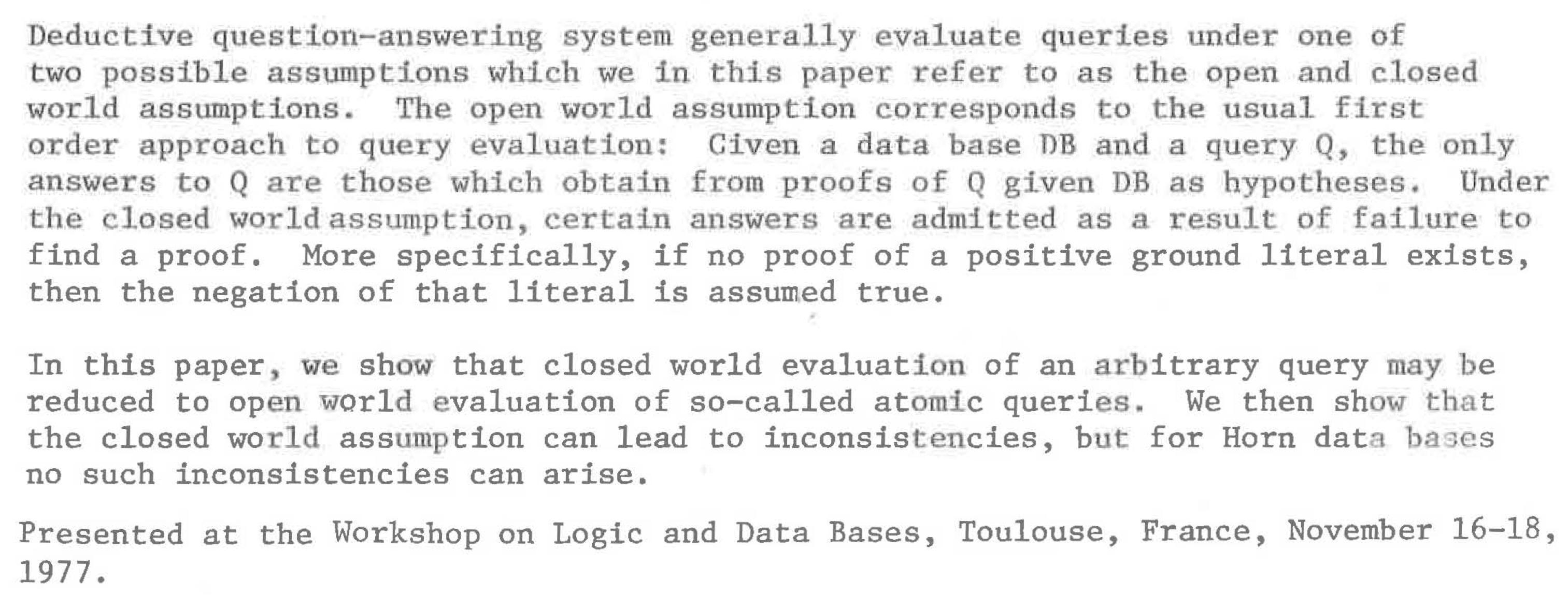}

Our second take on this ``controversy'' is that completions of relational databases by existential Horn clauses are Horn databases, and so for this reason we believe existential Horn clauses to be a better logic than description logic (which is not Horn) in the sense that we need not risk inconsistency in the closed to open world conversion process.  

Our third take on this ``controversy'' is that is closed-world reasoning per se isn't really that common, inasmuch as typical mathematics is done in open-world FOL.  The closed-world assumption is known in prolog/datalog as ``negation as failure''~\cite{negation}, where it is sometimes useful, and the language of existential Horn clauses doesn't even have negation.  Instead, the places where closed-world reasoning is needed, usually scenarios of the sort ``if there is no flight listed from USA to Belgium, then we can assume no flight exists'' are handled by ``bounded quantification'': rather than quantifying through {\it all the domain}, a universe so-large it is ``open'', we can quantify though {\it all flights in some existing table}, a universe so-small that it is a ``closed''.  Codd himself used bounded quantification through a database's ``active domain'' to prove his celebrated theorem~\cite{codd}, which itself can be thought of as implementing open-world FOL using ``closed-world'' (or rather, boundedly quantified) relational algebra.   

\end{itemize}
\subsubsection{SHACL/SHEX}
SHACL originated as an alternative to OWL for specifying the ``shapes'' that an RDF graph can take.  As such its default reasoning task is ``model checking'' ($\models$), unlike in OWL where the default reasoning task is ``entailment'' ($\vdash$).  SHACL's particular model checking semantics is often described as ``closed world'', but the exact relationship between DL and SHACL is not clear, and has been invested in e.g.~\cite{jan}.  We support SHACL's goal of ensuring data integrity, but note that its small scope (checking specific models, rather than proving entailment across all models) means that it isn't a panacea for the goals of the Semantic Web (a world with SHACL is better than a world without because data quality will be higher, but most challenges will still remain).  For the purposes of this paper, SHEX can be thought of as similar to SHACL, except that is uses regular expressions rather than shapes to describe its model conformance relation.

\section{Geospatial Semantic Web as Data integration case studies}
\label{case}
To help prove our argument, we have undertaken a number of case studies where we have formulated existing (geo)spatial semantic web problems as data migration and integration problems and demonstrated the benefits of an approached based on FOL and RL.  We have undertaken the case studies in the CQL tool~\cite{schultz_wisnesky_2017}, an open-source data integration tool based on FOL, RL, and a branch of math known as category theory~\cite{awodey}.  

\subsection{GML Manipulation}

The Open GIS GeographyMarkup Language (GML) Encoding Standard~\url{https://www.ogc.org/standards/gml} is a 437 page specification document describing an XML schema and intended semantics for spatio-temporal ontology. It models the world as a collection of ``cells'' representing geometry and topology to achieve a ``Multi-Layered Representation'' of a given scene applicable in different contexts but as such suffers from a ``multiplicity of representations'' problem: there are over 25 ways, for example, to represent a square in GML, and there is no recommended way to choose between them: \url{https://erouault.blogspot.com/2014/04/gml-madness.html}.  A variant of SPARQL that is sensitive to GML's intended semantics called GeoSPARQL exists which uses the "Well-known text" format to represent vector geometry. It  defines a SpatialObject concept that can be a Feature or a Geometry (or both) meant to inherit from by extending definitions of the General Feature Model, Simple Feature, and Geometry ontologies developed and standardized by the OGC as well. The defined relationships are binary properties relating two SpatialObject instances. It currently does not implement characteristics on properties where relevant and the definition of every property is mostly textual.

In our case study, our goal was to show how to mediate between different GML representations of the same shape using the language of regular logic; i.e., to show that existential Horn clauses can convert one kind of representation to another.  To achieve this we first converted GML from XML format to RDF format, and then we consider the RDF triples as a graph shallowly encoded in FOL.  For example, we started with:

\begin{footnotesize}
\begin{verbatim}
<gml:LinearRing><gml:posList>0 0 0 1 1 1 1 0 0 0</gml:posList></gml:LinearRing>
\end{verbatim}
and
\begin{verbatim}
<gml:LinearRing>
<gml:pos>0 0</gml:pos><gml:pos>0 1</gml:pos><gml:pos>1 1</gml:pos><gml:pos>1 0</gml:pos><gml:pos>0 0</gml:pos>
</gml:LinearRing>
\end{verbatim}
\end{footnotesize}
as two different GML representations of the unit square.  (We are eliding the XML within which the above fragments appear, for space reasons.)  Converted to relations, this XML becomes:
$$
\begin{tabular}{ | c | c |}
\multicolumn{2}{c}{{\sf gml:LinearRing}}  \\ 
 \hline ${\sf lr_1}$ & ${\sf pl_1}$  \\
 \hline 
\end{tabular}
\ \ \ \ 
\begin{tabular}{ | c | c |}
\multicolumn{2}{c}{{\sf gml:posList}}  \\ 
 \hline ${\sf pl1_1}$ & {\sf 0 0 0 1 1 1 1 0 0 0}  \\
 \hline 
\end{tabular}
\ \ \ \ \ \ \ \ 
\text{and}
\ \ \ \ \ \ \ \ 
\begin{tabular}{ | c | c |}
\multicolumn{2}{c}{{\sf gml:LinearRing}}  \\ 
 \hline ${\sf lr_1}$ & ${\sf pl_1}$  \\
 \hline 
\end{tabular}
\ \ \ \ 
\begin{tabular}{ | c | c |}
\multicolumn{2}{c}{{\sf gml:pos}}  \\ 
 \hline ${\sf pl1_1}$ & {\sf 0 0}  \\
\hline ${\sf pl1_1}$ & {\sf 0 1}  \\
\hline ${\sf pl1_1}$ & {\sf 1 0}  \\
\hline ${\sf pl1_1}$ & {\sf 1 1}  \\
 \hline 
\end{tabular}
$$
To convert from left to right, where we write ${\sf LinearRing}$ to indicate the left relation and ${\sf LinearRing}'$ to indicate the right relation of the same name, and where we assume the string concatenation function is present in our signature as function symbol $\oplus$ of arity two, we would write:
$$
\forall \ lr \ pl \ x_0 \ x_0 \ x_1 \ y_1 \ x_2 \ y_2 \ x_3 \ y_3, \ {\sf LinearRing}(lr,pl) \ \wedge \  {\sf posList}
(pl, x_0 \oplus y_0 \oplus x_1 \oplus y_1 \oplus x_2 \oplus y_2 \oplus x_3 \oplus y_3)
$$
$$
\to {\sf LinearRing'}(lr,pl) \ \wedge \ {\sf pos}
(pl, x_0 \oplus y_0) \ \wedge \ {\sf pos}
(pl, x_1 \oplus y_1) \ \wedge \ {\sf pos}
(pl, x_2 \oplus y_2) \ \wedge \ {\sf pos}
(pl, x_3 \oplus y_3) 
$$

To convert from right to left is harder, because we need to find four distinct points to make the square:
$$
\forall \ lr \ pl \ x_0 \ x_0 \ x_1 \ y_1 \ x_2 \ y_2 \ x_3 \ y_3, \
{\sf LinearRing'}(lr,pl) \ \wedge \ {\sf pos} (pl, x_0 \oplus y_0) \ \wedge \ {\sf pos} (pl, x_1 \oplus y_1) \ \wedge \ {\sf pos} (pl, x_2 \oplus y_2) \ \wedge \ {\sf pos} (pl, x_3 \oplus y_3)  
$$
$$
(x_0 \neq x_1 \wedge y_0 \neq y_1) \ \wedge \ (x_0 \neq x_2 \wedge y_0 \neq y_2) \ \wedge \ (x_0 \neq x_3 \wedge y_0 \neq y_3) \ \wedge \ (x_1 \neq x_2 \wedge y_1 \neq y_2) \ \wedge \ (x_1 \neq x_3 \wedge y_1 \neq y_3)  \ \wedge
$$
$$
(x_2 \neq x_3 \wedge y_2 \neq y_3) \ 
\to \ {\sf LinearRing}(lr,pl) \ \wedge \  {\sf posList}
(pl, x_0 \oplus y_0 \oplus x_1 \oplus y_1 \oplus x_2 \oplus y_2 \oplus x_3 \oplus y_3)
$$
Having the above relationships (``schema mappings'' in both directions~\cite{fagin}) in RL allows us to do more than simply convert one representation to another.  For example, we can define a normal form for squares: it is the schema that has {\it both} a position list and a list of positions, related according to the above (the two representations and the normal form are  ``Morita equivalent'', i.e., posses equivalent categories of models):
$$
\begin{tabular}{ | c | c |}
\multicolumn{2}{c}{{\sf gml:LinearRing}}  \\ 
 \hline ${\sf lr_1}$ & ${\sf pl_1}$  \\
 \hline 
\end{tabular}
\ \ \ \ 
\begin{tabular}{ | c | c |}
\multicolumn{2}{c}{{\sf gml:posList}}  \\ 
 \hline ${\sf pl1_1}$ & {\sf 0 0 0 1 1 1 1 0 0 0}  \\
 \hline 
\end{tabular}
\ \ \ \ 
\begin{tabular}{ | c | c |}
\multicolumn{2}{c}{{\sf gml:pos}}  \\ 
 \hline ${\sf pl1_1}$ & {\sf 0 0}  \\
\hline ${\sf pl1_1}$ & {\sf 0 1}  \\
\hline ${\sf pl1_1}$ & {\sf 1 0}  \\
\hline ${\sf pl1_1}$ & {\sf 1 1}  \\
 \hline 
\end{tabular}
$$

We can use the normal form to decide if two GML squares are the same according the above rules: we first convert two squares to this normal form (using the same chase procedure that converts them to each other's form) and we then compare the results-in-normal-form for equality (technically, isomorphism up to blank nodes), a technique from relational database theory typically used to find a canonical starting point for query optimization in a process called ``chase and backchase''~\cite{10.5555/932669}.  For example, the two original XML documents are equivalent because they both chase to the normal form above (up to blank node names) under the axioms of GML, such as they are.  In this way, we propose the use of libraries of regular logic axioms to be developed on top of formats such as GML, to foster interoperability in a world awash in multiplicity of representation.
      
\subsection{Combinatorial Maps instead of GML}

Like many geospatial systems, the intended semantics of GML, discussed above, is that of ``combinatorial maps''~\cite{comb}.  Astute readers will notice that the above GML manipulation scenario has more to do with the XML encoding of GML than GML's intended semantics, and in this section we directly axiomatize combinatorial maps in RL and propose the axiomatization as a starting point to constructing a fully formal semantics for all of GML.  Although it is more verbose than using function symbols, we propose a purely relational signature for GML, with the aim of maximum interoperability.

We define a relational signature for a combinatorial map of dimension three as having relation symbols $D$ of arity one, and $\beta_1, \beta_2, \beta_3$ of arity two.  Firstly, each $\beta_i$ must be a function on $D$:

$$
\forall x y, \beta_i(x,y) \to D(x) \wedge D(y) \ \ \ \ \forall xy, D(x) \to \exists y,  \beta_i(x,y)  \ \ \ \ \forall x y z, \beta_i(x,y) \wedge \beta_i(x,z) \to y = z 
$$
and each $\beta_i$ must also a permutation: that is, surjective and injective:
$$
\forall y, D(y) \to \exists x, \beta_i(x,y) \ \ \ \ \forall x y z, \beta_i(x,z) \wedge \beta_i(y,z) \to x = y
$$
and $\beta_2$ and $\beta_3$ must be involutions (must be equal to their own inverses):
$$
\forall xyz, \beta_2(x,y) \wedge \beta_2(y,z) \to x = z \ \ \ \
\forall xyz, \beta_3(x,y) \wedge \beta_3(y,z) \to x = z
$$
and $\beta_1 \circ \beta_3$ must be an involution:
$$
\forall vwxyz, \beta_3(x, w) \wedge \beta_1(w,y) \wedge  \beta_3(y, v) \wedge \beta_1(v, z) \to x = z \ \ \ \
$$
In GML proper, the maps can be $N$-dimensional and also ``partial'', elaborations on the above concept that are also FOL/RL-defineable.

Defining spatial concepts in this way allows for clean axiomatic definitions of a large class of typical (geo)spatial problems. For example the definition of a 3D parcel, a continuous real-world volume that is identified by a unique set of homogeneous property rights, is given in \cite{thompson2011axiomatic}.

\subsection{Graph Pushouts}

Geospatial information is often represented as graphs and one key operation operation on graphs that has not been discussed yet is the ``pushout'' operation~\cite{5521453}.  In this section we show how pushouts of graphs can be computed using the chase algorithm of existential horn clauses.  In this section, we are concerned with the scenario below:
$$
\includegraphics[width=4in]{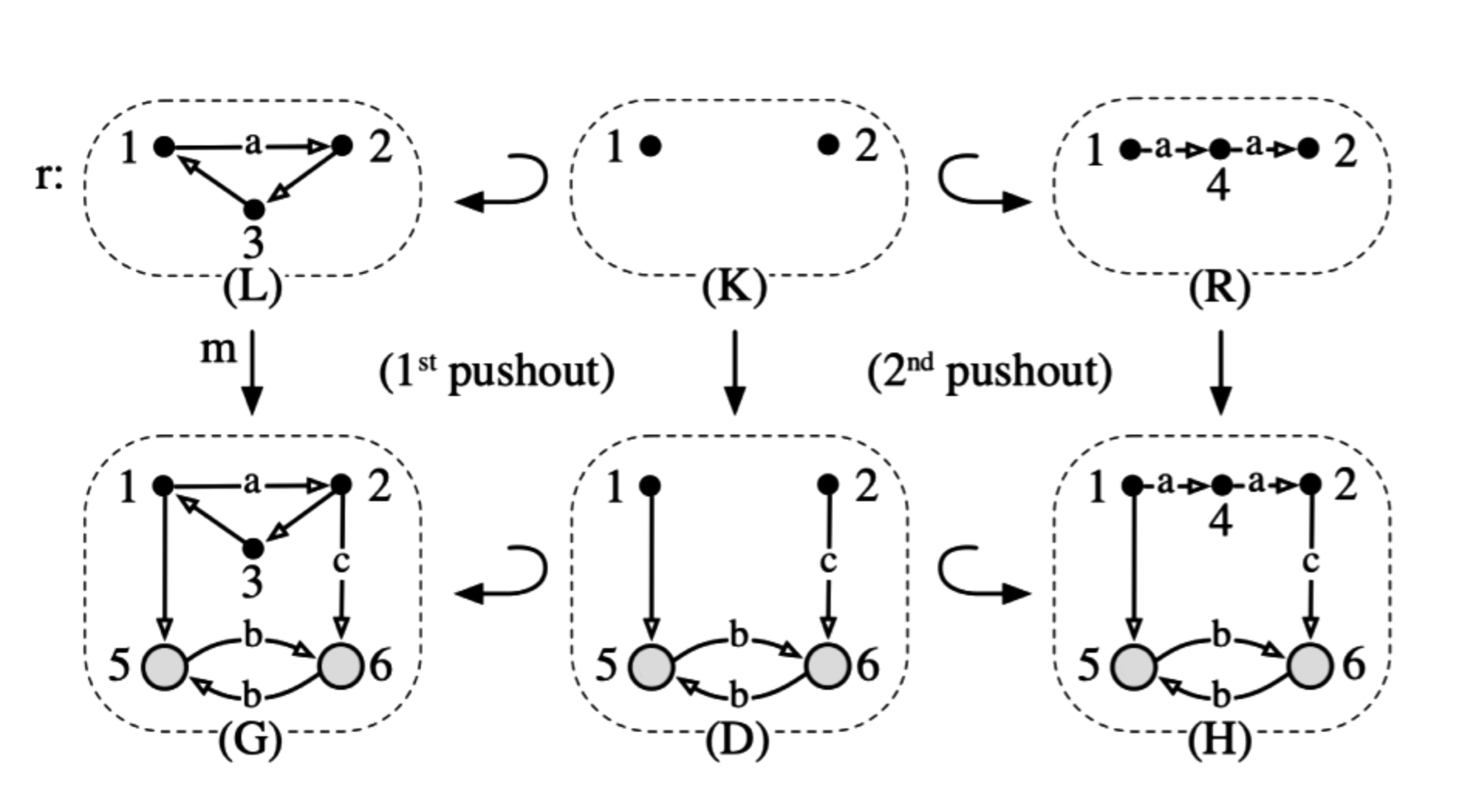}
$$
In other words we have relational databases and inclusions:

$$
K \ := \ {\sf node}({\sf n_1}), {\sf node}({\sf n_2}), {\sf nlabel}({\sf n_1}, {\sf 1}), {\sf nlabel}({\sf n_2}, {\sf 2})
$$

$$
L \ := \ K ,  {\sf node}({\sf n_3}), {\sf edge}({\sf e_1}), {\sf edge}({\sf e_2}), {\sf edge}({\sf e_3}),  {\sf nlabel}({\sf n_3}, {\sf 3}), {\sf elabel}({\sf e_1}, {\sf a}),  {\sf elabel}({\sf e_2}, -), {\sf elabel}({\sf e_3}, -),
$$
$$
{\sf src}({\sf e_1}, {\sf n_1}), {\sf dst}({\sf e_1}, {\sf n_2}),
{\sf src}({\sf e_2}, {\sf n_2}), {\sf dst}({\sf e_2}, {\sf n_3}),
{\sf src}({\sf e_3}, {\sf n_3}), {\sf dst}({\sf e_3}, {\sf n_1})
$$

$$
R \ := \ K ,  {\sf node}({\sf n_4}), {\sf edge}({\sf e_4}), {\sf edge}({\sf e_5}),   {\sf nlabel}({\sf n_4}, {\sf 4}), {\sf elabel}({\sf e_5}, {\sf a}),  {\sf elabel}({\sf e_5}, a), 
$$
$$
{\sf src}({\sf e_4}, {\sf n_1}), {\sf dst}({\sf e_4}, {\sf n_4}),
{\sf src}({\sf e_5}, {\sf n_4}), {\sf dst}({\sf e_5}, {\sf n_2})
$$

$$
X \ := {\sf node}({\sf n_5}), {\sf node}({\sf n_6}), {\sf nlabel}({\sf n_5}, {\sf 5}),
{\sf edge}({\sf e_6}), {\sf edge}({\sf e_7}),  {\sf edge}({\sf e_8}), {\sf edge}({\sf e_9}),
$$
$$
 {\sf elabel}({\sf e_6}, {\sf -}),  {\sf elabel}({\sf e_7}, c), {\sf elabel}({\sf e_8}, {\sf b}),  {\sf elabel}({\sf e_9}, b),
$$
$$
{\sf src}({\sf e_6}, {\sf n_1}), {\sf dst}({\sf e_6}, {\sf n_5}),
{\sf src}({\sf e_7}, {\sf n_2}), {\sf dst}({\sf e_7}, {\sf n_6}),
{\sf src}({\sf e_8}, {\sf n_5}), {\sf dst}({\sf e_8}, {\sf n_6}),
{\sf src}({\sf e_9}, {\sf n_6}), {\sf dst}({\sf e_9}, {\sf n_5})
$$

$$
D \ := \ K , X \ \ \ \ G \ := \ L, X \ \ \ \ H := \ R, X
$$
 The example above is contrived (by the original authors of~\cite{5521453}) in the sense that the relationships between the databases are entirely that of inclusion.
In a more general case the node and edge identifiers would be different between each graph, and the relationships between graphs would be ``homomorphisms'', not inclusions. In that case, we cannot simply use union to compute the pushout of three graphs.  Instead, we have to use the chase algorithm: after disjointly unioning together $R$ and $D$, we must ``merge'' (technically, quotient) the result to take into account their overlap $K$.  Let $h^{K,D}$ and $h^{K,R}$ be binary relations corresponding to the given morphisms $K \to D$ and $K \to R$, respectively.  The required formulae are simply: $\forall x y, h^{K,D}(x, y) \to x = y$ and $\forall x y, h^{K,R}(x, y) \to x = y$, that is, nodes, edges, and labels should be equated in the pushout graph exactly when they are the source and target of the input overlap morphism.  


\subsection{Geographic Partition/Lattice Merge}

One of the advantages to working in FOL and RL is the many connections between those logic and various areas of algebra, for example, lattice theory; the reason this connection exists is because many algebraic structures are defined fragments of FOL such as RL.  Such is the case for lattice theory specifically, which means that we can use the chase algorithm to merge geographical information that is structured according to lattices, such as the classifications of areas into e.g. arid, semi-arid, wetland, marsh, ocean, etc, where the lattice relationship tracks e.g. ``is a sub-classification of'', so in some ways this case study is a blueprint for how to avoid reasoning about isA relationships in OWL when dealing with partitions of a space.  The two inputs to our case study our shown below; they are two partitionings of the same space, along with two taxonomies and an assignment from each partitioning to its corresponding taxonomy, show at left and right:

\includegraphics[width=2.8in]{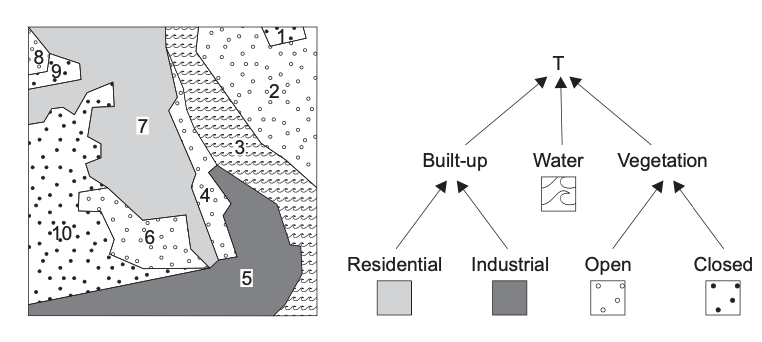}\hspace{.6in}
\includegraphics[width=2.8in]{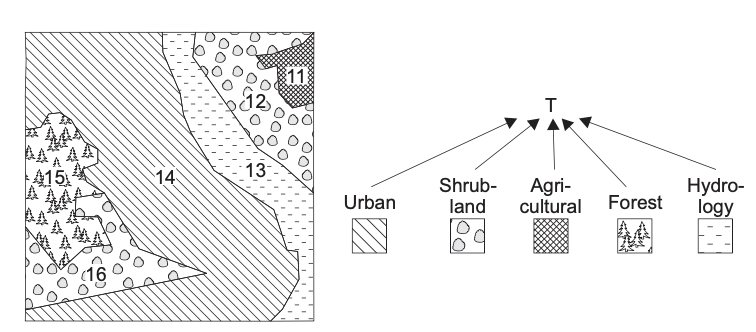}

\noindent
Let us write $X$ for the set of points in our geographic region, and $\sim_1$ and $\sim_2$ for the two partitionings thereof, and $\preceq_1$ and $\preceq_2$ for the two join semi-lattice relations (where $A \to B$ in the diagram means $A \prec B$), and $g_1$ and $g_2$ for the function associating each partition to its label according to the corresponding visual shading.  Rather than work with the $\preceq_i$ predicates we prefer to work with the least-upper-bound function of join semi-lattices, written ${\sf lub}_i$.  We have as usual $x \preceq_i y$ iff $x = {\sf lub}_i(x, y)$:
$$
\textsf{Built-up} \preceq_1 \top_1  \ \ \ \ {\sf Residential}  \preceq_1 \textsf{Built-up} \ \ \ \  \ldots 
$$
$$
\top_1 = {\sf lub}_1(\top_1, \textsf{Built-up}) \ \ \ \  \textsf{Built-up} = {\sf lub}_1(\textsf{Built-up}, {\sf Residential}) \ \ \ \ \ldots
$$
For simplicity, in this section we will use multi-sorted FOL, and assume four sorts: the sort of points of our geographic region, written $X$, and three sorts of labels, $T_0$ (output), $T_1$ (input) and $T_2$ (input).  The input to our case study is thus expressed in FOL as:

\begin{itemize}
\item two equivalence relations on (partitionings of) $X$, written $\sim_1$ and $\sim_2$; equivalence relations are reflexive, symmetric, and transitive:
$$
\forall x:X, \ x \sim_i x  \ \ \ \ \forall x y : X, \ x \sim_i y \to y \sim_i x \ \ \ \ 
\forall x y z : X, \ x \sim_i y \wedge y \sim_i z \to x \sim_i z
$$
\item two join semi-lattices, one on $T_1$ and one on $T_2$, written ${\sf lub}_1$ and ${\sf lub}_2$, each (necessarily) with a maximal (absorbing) element.  A join semi-lattice is a function:
$$
\forall x y :T_i, \ \exists y:T_i, \ {\sf lub}_i(x, y, z)
\ \ \ \ 
\forall x y z z' : t_i, {\sf lub}_i(x, y, z) \wedge {\sf lub}_i(x, y, z') \to z = z'
$$
that is commutative, idempotent, and associative:
$$
\forall x : T_i, \ {\sf lub}_i(x,x,x) \ \ \ \ 
\forall x y z : T_i, \ {\sf lub}_i(x,y,z) \to {\sf lub}_i(y,x,z)   
$$
$$\forall u v w x y y' z : T_i, \  
{\sf lub}_i(v,w,x) \wedge {\sf lub}_i(x,z,y) \wedge
{\sf lub}_i(w,z,u) \wedge {\sf lub}_i(v,u,y')
\to y = y' 
$$
and with a maximal (absorbing) element, $\top_i$:
$$
\forall x:T_i, \ {\sf lub}_i(x, \top_i) = \top_i
$$
Non-empty finite semi-lattices always have maximal elements, as an easy inductive argument shows.

\item two functions, written $g_1 : X \to T_1$ and $g_2 : X \to T_2$:
$$
\forall x:X, \ \exists y:T_i, \ g_i(x,y) \ \ \ \ \forall x y : X, \ \forall z : T_i, \ g_i(x,y) \wedge g(x,z) \to y=z
$$
each of which respects the associated equivalence relation:
$$
\forall x x' : X, \forall y y' : T_i, \ g_i(x, y) \wedge g_i(x', y') \wedge x \sim_i x' \to y = y'
$$
\end{itemize}
The output will be an equivalence relation $\sim_0$ on $X$, and a lattice on $T_0$, written (${\sf lub_0}$, ${\sf glb_0}$, $\top_0$, $\bot_0$), and a function, $g_0 : X \to T_0$, that respects $\sim_0$.  Constructing the output from the two inputs is conceptually a two step chase process: first, we compute the intersection partitioning $\sim_1 \cap \sim_2$ of the space $X$; this is $\sim_0$:
$$
\forall x y : X, \ x \sim_1 y \wedge x \sim_2 y \to \ x \sim_0 y   
$$
The next step computes ${\sf lub_0}, {\sf glb_0}, \top_0, \bot_0$ and  involves auxiliary sort-conversion functions $\tau_1 : T_1 \to T_0$ and $\tau_2 : T_2 \to T_0$, with $i=1,2$ below:
$$
\forall x : T_i, \ \exists y : T_0, \ \tau_i(x, y) \ \ \ \ \forall x y z z' : T_i, \ \tau_i(x, y, z) \wedge \tau_i(x, y, z') \to z = z'
$$
These functions are used to populate ${\sf lub}_0$:
$$
 \forall x y : T_i, \ \forall x' y' : T_0, \ {\sf lub}_i(x, y, z) \wedge \tau_i(x,x') \wedge \tau_i(y,y') \wedge \tau_(z,z') \to {\sf lub}_0(x', y', z') 
$$
Besides the join semi-lattice axioms, we require the axioms for a meet semi-lattice, which is a function ${\sf glb_0}$:
$$
\forall x y :T_i, \ \exists y:T_i, \ {\sf lub}_i(x, y, z)
\ \ \ \ 
\forall x y z z' : t_i, {\sf lub}_i(x, y, z) \wedge {\sf lub}_i(x, y, z') \to z = z'
$$
that is idempotent, associative, and commutatitve, with a bottom (identity) element:
$$
\forall x : T_0, \ {\sf glb}_0(x, \bot_i) = x \ \ \ \  
\forall x : T_0, \ {\sf glb}_0(x,x,x) \ \ \ \ 
\forall x y z : T_0, \ {\sf glb}_0(x,y,z) \to {\sf glb}_0(y,x,z)   
$$
$$\forall u v w x y y' z : T_0, \  
{\sf glb}_0(v,w,x) \wedge {\sf glb}_0(x,z,y) \wedge
{\sf glb}_0(w,z,u) \wedge {\sf glb}_0(v,u,y')
\to y = y' 
$$
as well as interchange-laws stating the ${\sf glb_0}$ and ${\sf lub_0}$ define the same partial order:
$$
{\sf glb}_0(x, y, z) \wedge {\sf lub_0}(x, z, w) \to x = w
\ \ \ \ 
{\sf lub}_0(x, y, z) \wedge {\sf glb}_0(x, z, w) \to x = w
$$
Finally, we compute $g$ (which already has function axioms) using greatest upper bounds:
$$
 \forall w : T_i, \ \forall x y z : T_0, \  \tau_1(w,x) \wedge \tau_2(w,y) \wedge {\sf glb_0}(x, y, z) \to  g(w, z)
$$

The above process, of completing a join-semilattice into an actual lattice by adding greatest lower bounds as needed, is known as ``Dedekind-MacNeille'' completion ~\cite{lattice}, and results in output taxonomy labels that are not present in either input.  The output partitioning and taxonomy is shown below, obtained with the chase process as usual. 
$$
\includegraphics[width=2.4in]{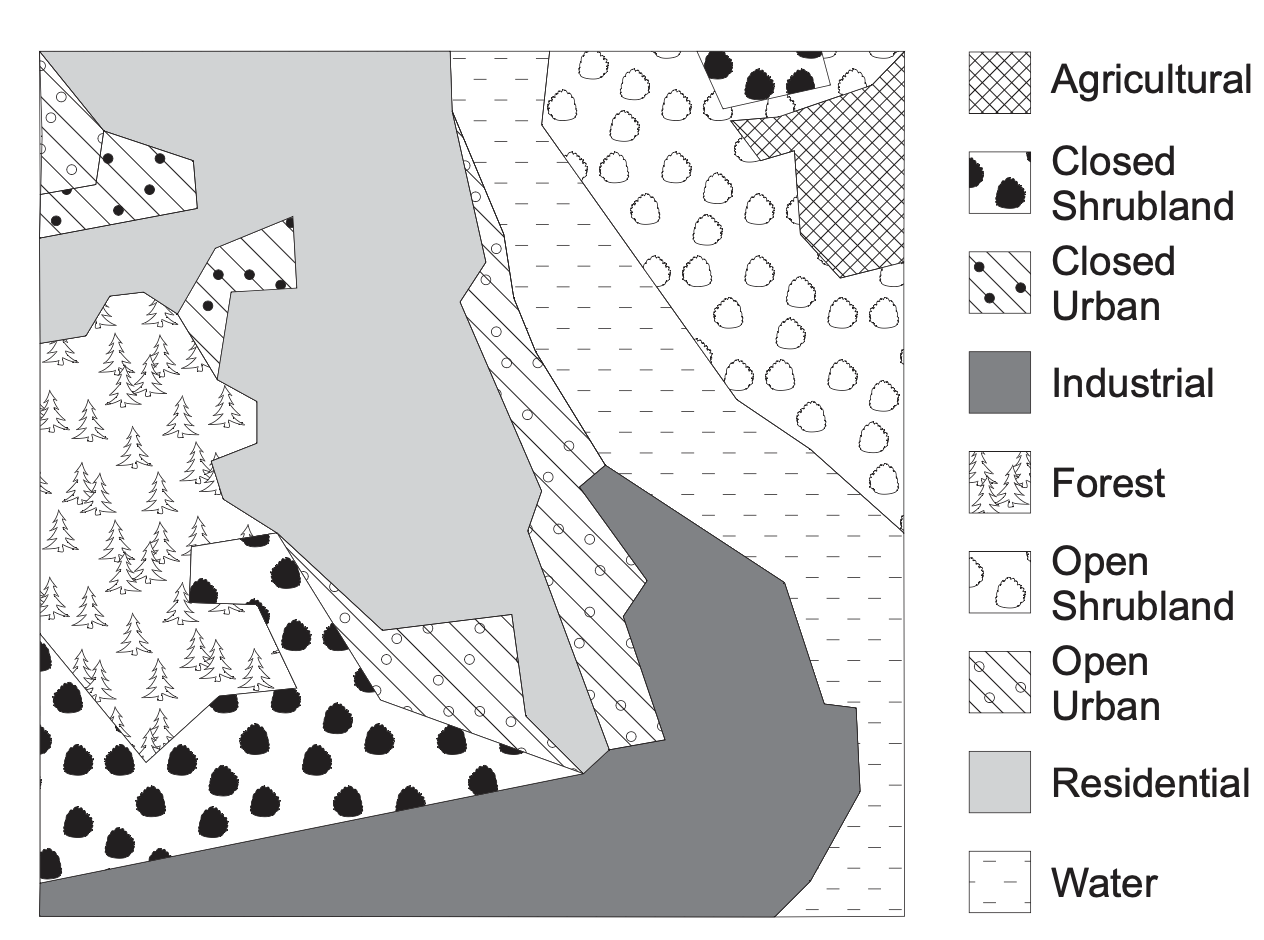}
\hspace{.4in}
\includegraphics[width=2in]{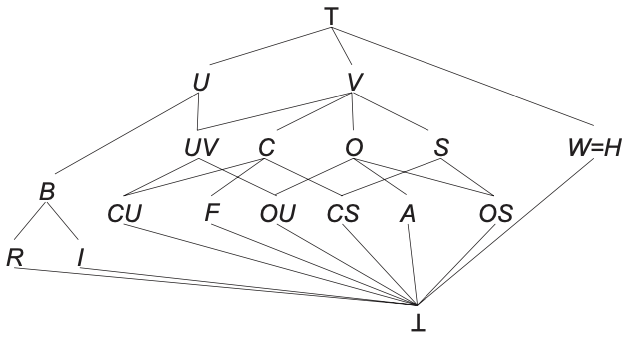}
$$
Variations on the integration of structured semantics is studied under ``algebraic data integration''~\cite{schultz_wisnesky_2017}.

\subsection{Computing Partitions from Polygons}
Left unspecified in the above is how to obtain the input partitionings from, say, a set of polygons.  In its full generality this problem may require algorithms like that even-odd algorithm (\url{https://en.wikipedia.org/wiki/Even–odd_rule}); in this section, we describe how RL can compute partitionings from polygons given as lists of points, such as in our GML example.   For expediency, we will use function symbols and abbreviate $f(x)$ as $fx$ and use infix notation in places.  We note that this example uses equational logic, itself a fragment of RL, and so can easily be checked in e.g. C or java. 

We require additional sorts, $\texttt{String}$, $\texttt{B}$, $\texttt{Int}$, $\texttt{PL}$, $\texttt{E}$, and $\texttt{EL}$, and recall that $X$ is the sort of points.  We require function symbols:
\begin{small}
\begin{verbatim}
Point : Int,Int -> X       PointListNil : PL       PointListCons : X ,PL -> PL

Edge : Int,Int,Int,Int -> E      EdgeListNil : EL        EdgeListCons : E,EL -> EL

True : B        False : B       && : B,B -> B      >= : Int,Int -> B

stringToPointList : String -> PL
\end{verbatim}
\end{small}
We will omit axiomatizating Booleans (easy) and Integers (tedious) and converting Strings to lists of points (very tedious).  We thus have:
\begin{small}
\begin{verbatim}
pointListToEdgeList (PointListCons (Point a b) (PointListCons (Point c d) ps))
  = EdgeListCons (Edge a b c d) (pointListToEdgeList (PointListCons (Point c d) ps))
pointListToEdgeList (PointListCons x PointListNil) = EdgeListNil
pointListToEdgeList PointListNil = EdgeListNil

crossProduct (Edge x1 y1 x2 y2) = (x1*y2) - (y1*x2)

vectors (Edge x1 y1 x2 y2) (Point px py) = Edge (x2-x1) (y2-y1) (px-x1) (py-y1)              

pointInEdgeList p EdgeListNil = True
pointInEdgeList p (EdgeListCons edge y) = crossProduct (vectors edge p) >= 0 && pointInEdgeList p y

pointInPointList p pl = pointInEdgeList p (pointListToEdgeList pl)
\end{verbatim}
\end{small}
For example, we have:
\begin{small}
\begin{verbatim}
poly = PointListCons (Point 0  0) (PointListCons (Point 10 0) (PointListCons (Point 10 10) 
      (PointListCons (Point 0 10) (PointListCons (Point  0 0)  PointListNil))))
poly = stringToPointList "0 0 10 0 10 10 0 10 0 0"
True = pointInPointList (Point 5 5) poly
False = pointInPointList (Point 15 15) poly
\end{verbatim}
\end{small}
For each linear ring ${\sf lr}$, we enumerate all points in the plane, populating $\sim_{\sf lr}$ (two points are in the same partition if and only if they are both inside or outside of the linear ring):
\begin{small}
\begin{verbatim}
forall s pl : String, x y x' y' : X, LinearRing(lr, pl) /\ posList(pl, s) /\
 pointInPointList (Point x y) (stringToPointList s) /\ pointInPointList (Point x' y') (stringToPointList s) -> 
(Point x y) ~_lr (Point x' y')
\end{verbatim}
\end{small}
Note that many choices of partitioning are possible; the above algorithm works only for convex polygons and is not fine-grained as the even-odd algorith; however, it is the smallest algorithm we are aware of.  

\section{A Proposal For A More Reasonable Semantic Web}
\label{proposal}
To extend the semantic web we embrace JSON-LD-LOGIC by T. Tammet and G. Sutcliffe so that we may use FOL reasoners such as E~\url{http://www.eprover.org} and Vampire~\cite{vampire} and Z3~\cite{z3} and dozens of chase engines~\cite{10.1145/3034786.3034796}.  In such a world, many of the frictions of using ``RDF'' would not exist in the first place.  We now elaborate on the advantages to our proposal before describing TPTP in more detail.

\begin{itemize}

\item {\it Future-proof.}  There are deep philosophical and computational reasons why FOL and RL have emerged as  dominant logics over the past hundreds of years, and these reasons will be just as true in the future as they are today (i.e., the reasons are about computation and knowledge, not particular software systems or file formats.)  At a sociological level, using FOL and RL is future-proof because these communities are lively and growing.

\item {\it SQL and Datalog interoperability.}  SQL has been the dominant data manipulation for many decades, and owing to its canonical roots as a fragment for FOL, is likely to remain so for a long time.  Codd's theorem~\cite{codd} provides an easy way to translate between FOL and SQL.  As an added bonus, FOL and RL also have strong connections to Datalog and Prolog; the TPTP format, described later, originated as a prolog format.

\item {\it Automated Theorem Proving.}The automated theorem proving community and vendors have produced hundreds of first-order provers and model checkers, including E~\url{http://eprover.org}, Vampire~\cite{vampire}, and Z3~\cite{z3}.  Compatibility with the TPTP language ensures this ecosystem can be leveraged.  

\item {\it Backward Compatible with DL and parts of RDF/OWL.}  Although there are fragments of RDF and OWL that do not have direct have meaning in FOL (such as the paradoxical triple $X(X,X)$ for a URL $X$), there are fragments of them that do have meaning in FOL.  Moreover, DL is a fragment of FOL.  This means that we would not lose much existing work on the semantic web.


\item {\it Tooling support for data migration/integration.}  Because mainstream work on data integration and migration is done with FOL and RL, our proposal would allow dozens of chase engines~\cite{10.1145/3034786.3034796} and first-order reasoners such as Vampire~\cite{vampire}.  A previous attempt to connect the semantic web to first-order logic and TPTP format in particular was called ``inference web'' and was focused on providing explainability for semantic web applications~\cite{DBLP:conf/cade/SilvaSCDRM08}.  This project and others have built OWL to FOL translators. 

\item {\it Theoretical support for spatio-temporal reasoning.}  As our case studies demonstrate, FOL and RL are sufficient to capture many aspects of spatio-temporal reasoning.  Additionally, FOL can be extended to deal with time and space in many rigorous ways, for example~\cite{ttt}.

\end{itemize}

\subsection{Logic formats}

TPTP, \url{https://www.tptp.org/}, is a text format used, among other places, in an annual FOL theorem proving competition.  In TPTP, $\forall xy, P$ is written \texttt{![x,y]:P} and $\exists xy, P$ is written \texttt{?[x,y]:P}; also, $\neg$ is written $\sim$ and $\wedge$ as \texttt{\&} and $\vee$ as $\texttt{|}$ and $\to$ as $\texttt{=>}$; the signature is implicit.  In its simplest form, a TPTP file is simply a list of named axioms, where each is marked with the fragment of FOL it is supposed to live in (``fof'' being the top level ``first-order form'', i.e., all of FOL).  Variables are written uppercase and idenfiers in lowercase.  For example:

\begin{small}
\begin{verbatim}
fof(all_created_equal, axiom, (![H1,H2] : ((human(H1) & human(H2)) => created_equal(H1,H2)))). 

fof(john, axiom, (human(john))). 

fof(john_failed, axiom, (grade(john) = f)). 

fof(someone_got_an_a, axiom, (?[H] : ( human(H) & grade(H) = a ) )). 

fof(distinct_grades, axiom, (a != f )). 

fof(grades_not_human, axiom, (![G] : ~human(grade(G)))). 

fof(someone_not_john, axiom, (?[H] : (human(H) & H != john ))). 
\end{verbatim}      
\end{small}

As remarked in the definitions section, axioms such as the above are sufficient to include relational databases and graph databases and more, with existential quantification used to encode blank nodes / labelled nulls as usual.  However, there are extensions to TPTP that treat models as a first-class concept and also allow to encode proof derivations~\cite{tptp2}.  We note that our critique of URIs in RDF applies to TPTP as well: we would prefer if axioms were not required to have names.  Indeed, TPTP is far from perfect: its multi-sorted semantics, which we will see next, does not allow empty sorts, an unfortunate omission that adds verbosity when doing database theory in particular using FOL.

\subsection{Multi-sorted TPTP}

The TPTP format includes many variations of FOL, including multi-sorted FOL.  In this section, we express some of the lattice merge case study using multi-sorted TPTP, which requires explicit signatures.  The sort (type) of booleans is written \texttt{\$o} and the ``type of types'' is written \texttt{\$tType}, and arities are written using \texttt{>}:
\begin{small}
\begin{verbatim}
tff(X_type, type, x: $tType).
tff(T0_type, type, t0: $tType). tff(T1_type, type, t1: $tType). tff(T2_type, type, t2: $tType).

tff(equiv0_type, type, equiv: (x*x) > $o). 
tff(equiv1_type, type, equiv: (x*x) > $o). 
tff(equiv2_type, type, equiv: (x*x) > $o). 

tff(isect_part, axiom, ! [X : x, Y : x] : (equiv1(X,Y) & equiv2(X,Y)) => equiv0(X,Y)).  

tff(lub0_type, type, lub0: (t0*t0*t0) > $o).
tff(lub0_comm , axiom, ! [ X : t0, Y : t0, Z : t0 ] : lub0(X, X, X)).
tff(lub0_idem , axiom, ! [ X : t0 ] : (lub0(X,Y,Z) & lub0(Y,X,Z)).
etc
\end{verbatim}
\end{small}

Multi-sorted TPTP also includes optional definitions for numerals and arithmetic, and allows functions, not just functional relations to be expressed~\cite{tff}.  

\subsection{JSON-LD-Logic}

Because of TPTP's straightforward syntax and semantics, it is easy to encode in other formats, including structures such as LISP / Web Assembly's S-Expressions. In this section, we briefly describe JSON-LD-LOGIC~\cite{tptpjson}, a specification compatible with both JSON-LD (\url{https://json-ld.org}) and TPTP format, allowing interchange between the two.  And because JSON-LD is aware of RDF and OWL, it is possible to intermix FOL and RDF/OWL concepts.  For example, here we define a previously described GML square type that has both a position list and a list of positions.

\begin{small}


\begin{verbatim}
[
 {"@context": { "@vocab":"http://foo.org/"}, "@id":"pete", "@type": "male",
  "father":"john", "son": ["mark","michael"],
  "@logic": [
   "forall",["X","Y"],[[{"@id":"X","son":"Y"},"&",{"@id":"X","@type":"male"}],"=>",{"@id":"Y","father":"X"}]
  ]},   

 ["if", {"@id":"?:X","http://foo.org/father":"?:Y"},  {"@id":"?:Y","http://foo.org/father":"?:Z"}, 
  "then", {"@id":"?:X","grandfather":"?:Z"}]
]
\end{verbatim}
\end{small}
Becomes, where \texttt{\$arc} is the name of the relation of RDF triples in JSON-LD:
\begin{small}
\begin{verbatim}
fof(frm_1,axiom, ((! [X,Y] : 
     (($arc(X,'http://foo.org/son',Y) & $arc(X,'http://www.w3.org/1999/02/22-rdf-syntax-ns#type',male)) 
      => $arc(Y,'http://foo.org/father',X))) & ($arc(pete,'http://foo.org/son',michael) & 
         ($arc(pete,'http://foo.org/son',mark) & ($arc(pete,'http://foo.org/father',john) & 
          $arc(pete,'http://www.w3.org/1999/02/22-rdf-syntax-ns#type',male)))))).

fof(frm_2,axiom, (! [X,Y,Z] : (($arc(Y,'http://foo.org/father',Z) & $arc(X,'http://foo.org/father',Y)) 
      => $arc(X,grandfather,Z)))).
\end{verbatim}
\end{small}
There's also an encoding into XML:

\begin{small}
\begin{verbatim}
fof(the_name, axiom, ?[X] : (p(X,1) => q(X,2))). 
\end{verbatim}
\end{small}
becomes:
\begin{small}
\begin{verbatim}
<scl:formula  xmlns:scl="http://scluripart">
  <scl:quantifier scl:name="exists"  scl:variable="x">
    <scl:connective scl:name="implies">
      <scl:predicate scl:name="p"><scl:term scl:name="x"/><scl:term scl:name="1"/></scl:predicate>  
      <scl:predicate scl:name="q"><scl:term scl:name="x"/><scl:term scl:name="2"/></scl:predicate>        
    </scl:connective>                      
  </scl:quantifier>
</scl:formula>
\end{verbatim}
\end{small}%

\subsection{Conclusion}
\label{conc}

We have described how to accelerate these goals by reconsidering the semantic web’s foundations: we have argued that building the semantic web should be approached as a traditional data migration and integration problem~\cite{fagin} at a massive scale, so that a huge amount of existing tools and theories~\cite{tptp1} can be deployed to the semantic web's benefit.  We have proposed to embrace Json-LD-Logic as bridge from DL to RL and provided case studies and theoretical arguments in support of our argument.

\bibliographystyle{plain} 
\bibliography{bib}

\end{document}